\documentclass[conference]{IEEEtran}
\IEEEoverridecommandlockouts
\usepackage{cite}
\usepackage{amsmath,amssymb,amsfonts}
\usepackage{algorithmic}
\usepackage{graphicx}
\usepackage{textcomp}
\usepackage{xcolor}
\usepackage{algorithm}
\ifCLASSOPTIONcompsoc
\usepackage[caption=false,font=normalsize,labelfon
t=sf,textfont=sf]{subfig}
\else
\usepackage[caption=false,font=footnotesize]{subfig}
\fi
\graphicspath{{./figs/}}
\def\BibTeX{{\rm B\kern-.05em{\sc i\kern-.025em b}\kern-.08em
    T\kern-.1667em\lower.7ex\hbox{E}\kern-.125emX}}

\newtheorem{theorem}{Theorem}
\begin{document}

\title{Probabilistic Bounds on the End-to-End \\ Delay of Service Function Chains \\ using Deep MDN\\
}

\author{\IEEEauthorblockN{Majid Raeis, Ali Tizghadam and Alberto Leon-Garcia}
\IEEEauthorblockA{\textit{Department of Electrical and Computer Engineering} \\
\textit{University of Toronto}, Toronto, Canada \\
Emails: m.raeis@mail.utoronto.ca, ali.tizghadam@utoronto.ca  and alberto.leongarcia@utoronto.ca}
}

\maketitle

\begin{abstract}
Ensuring the conformance of a service system’s end-to-end delay to service level agreement (SLA) constraints is a challenging task that requires statistical measures beyond the average delay.  In this paper, we study the real-time prediction of the end-to-end delay distribution in systems with composite services such as service function chains. In order to have a general framework, we use queueing theory to model service systems, while also adopting a statistical learning approach to avoid the limitations of queueing-theoretic methods such as stationarity assumptions or other approximations that are often used to make the analysis mathematically tractable. Specifically, we use deep mixture density networks (MDN) to predict the end-to-end distribution of the delay given the network's state. As a result, our method is sufficiently general to be applied in different contexts and applications. Our evaluations show a good match between the learned distributions and the simulations, which suggest that the proposed method is a good candidate for providing probabilistic bounds on the end-to-end delay of more complex systems where simulations or theoretical methods are not applicable.
\end{abstract}


\section{introduction}\label{intro}
Measuring the quality of service (QoS) for a particular service or application is often a challenging task. In addition, most applications and services are composed of more fine-grained services, making quality of service assessment even more complicated. Service function chaining is one such example in which the network services consist of an abstract sequence of service functions (SFs)~\cite{sfc}, where each service function provides a specific service, such as load balancing, deep packet inspection, etc. Therefore, the end-to-end performance of the service chain depends on the performance of the constituent SFs, as well as the order of the SFs that are visited by the incoming packets. The end-to-end delay of a service chain is one of the important measures of the QoS, particularly when providing time-sensitive services in which the packets must be processed within some specific deadline. In this case, real-time prediction of the delay \emph{distribution} can be much more informative compared to the existing single-value delay prediction methods. For instance, the predicted distribution can be used for designing an \emph{admission controller} that rejects the packets that have a high probability of missing the deadline. Moreover, the predicted distribution can be used for other control purposes such as auto-scaling of virtual network functions (VNFs) in a service chain. We refer the reader to~\cite{survey, auto_scal, admission} for some existing works on the performance, auto-scaling and admission control of VNF chains. 

In order to study the end-to-end performance of such systems, we take a general approach and do not limit ourselves to a specific application. In particular, we use queueing theory as a general framework for modeling service systems.
However, we do not adopt the traditional queueing theoretic methods because of their limitations and unrealistic assumptions. Instead, we take a statistical learning approach, without making any assumptions about the network topology, or service and inter-arrival time distributions. Specifically, we study the end-to-end delay of a service network, which is one of the important measures of the quality of service, by applying queueing models to the underlying services (SFs in the context of SFC), so that the theoretical analysis is replaced with statistical learning methods. It should be noted that \emph{real-time} prediction and analysis of the end-to-end delay in queueing networks, under \emph{non-stationarity} assumptions, is an \emph{under-explored} problem~\cite{i2018}, which will be studied in this paper.

\subsection{Background and Previous Work}
Here, we briefly review the literature on \emph{real-time} delay prediction and analysis of queueing systems, as opposed to \emph{steady state} studies. In order to perform real-time prediction, different types of information such as queue length and delay history are typically used. We classify delay analysis in service systems based on two aspects: the analysis technique (queueing-theoretical versus data-based) and the system topology (single-stage versus multi-stage network). Let us first begin with the queueing-theoretic methods for the single-stage queueing systems.

One of the earliest work on predicting a customer's waiting time in a multi-server queueing system is~\cite{whitt99}. This paper investigates the possibility of improving delay predictions by exploiting information about the system state, and the elapsed service time of the customers in service, under non-exponential service time assumptions. Following on~\cite{whitt99}, the performance of alternative queue-length-based and delay-history-based predictors for multi-server queues have been studied in \cite{ibrahim_thesis}. 

In contrast to the single-stage case, real-time delay prediction in \emph{multi-stage} queueing systems has not yet been extensively studied. One of the few examples in this category, which is also closely related to this paper, is the approximation model proposed in \cite{gue} for predicting the sojourn-time distribution of the customers in multi-stage systems. Using phase-type distributions, the authors develop a model for approximating sojourn-time distributions based on queue length information.  
Although the authors in \cite{gue} use general inter-arrival and service time distributions, the method assumes heavy traffic, stationary distributions and knowledge of the system parameters and the network topology.

Limitations of the queueing-theoretic analysis have led to recent interest in data-based methods such as machine-learning algorithms and data-mining techniques. 
Combining process mining and queueing-theoretic results, a technique called queue-mining is introduced in~\cite{sender} for predicting waiting times in service systems.  In~\cite{ang}, the authors propose a new predictor, called Q-Lasso, which combines the Lasso method from statistical learning and fluid models from the queueing theory. Similar to \cite{sender} and~\cite{ang}, most of the existing works in this area focus on single-value delay predictions and provide no information on the distribution of the delay. A closely related work to this paper is~\cite{majid}, which studies delay distribution prediction in single stage queueing systems using delay history information. Taking a statistical learning approach, the method in \cite{majid} is capable of predicting the conditional distribution of the delay under non-stationary conditions, without any knowledge of the system parameters.

\subsection{Motivation}
Both queueing-theoretic as well as data-based methods have their own advantages and shortcomings. One of the main disadvantages of queueing-theoretic methods is that the analysis can easily become intractable when introducing more realistic assumptions, such as general non-stationary inter-arrival and service time distributions. Furthermore, the queueing-theoretic methods require knowledge of the model parameters and the network topology, which might not be available and need to be estimated as well. Another shortcoming of the these methods is their limitation in exploiting all the available information. On the other hand, the prediction method and the feature selection process in data-based predictions are usually specialized for a particular application and do not provide much insight into the behaviour of general queueing networks. Uncertainty of the estimations and the distribution of the waiting times are additional pieces that are often missing in data-based methods.
The combination of these reasons motivated us to use statistical learning methods to study queueing models under more realistic assumptions, such as \emph{non-stationary} arrivals and \emph{non-exponential} service times. Furthermore, our proposed method enables us to estimate the \emph{conditional distribution} of the end-to-end delay, which is much more informative than single-value predictions. 
 
The remainder of this paper is organized as follows. In Section~\ref{sys_model}, we describe the queueing system model and formulate the problems that we study in this paper. We begin our analysis by providing some theoretical results on the end-to-end delay of the service networks in Section~\ref{analysis}. In Section~\ref{GMM}, we propose to use Gaussian mixture models as an approximation of the delay distribution, parameters of which can be estimated using mixture density networks. The evaluation of the proposed methods are presented in Section~\ref{eval}. Finally, Section~\ref{con} presents the conclusions.

\section{System Model and Problem Setting}\label{sys_model}
\begin{figure}[!t] 
\centering
\subfloat[]{\includegraphics[scale=0.55]{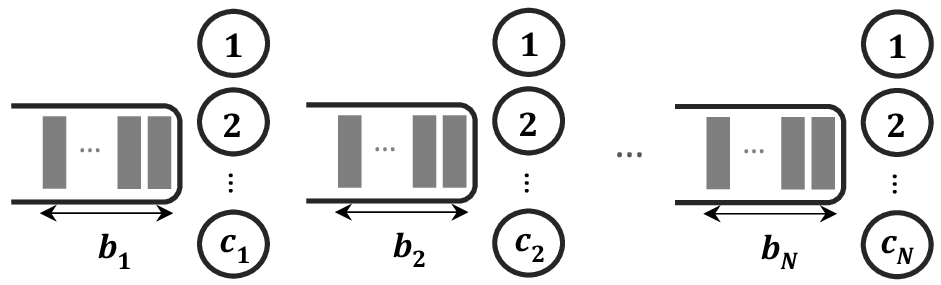}
\label{fig:tandem_topo}}
\hfill
\subfloat[]{\includegraphics[scale=0.55]{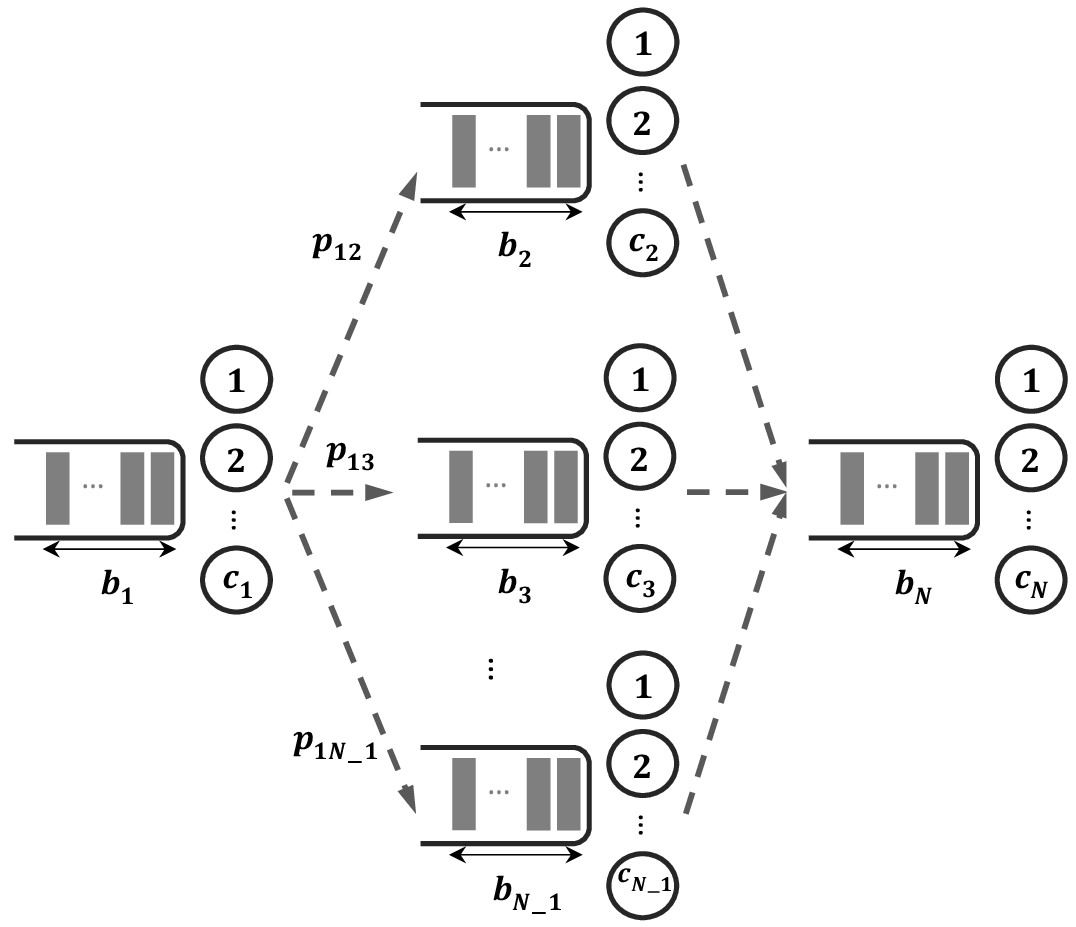}
\label{fig:acyclic_topo}}
\caption{Network topologies (a) Tandem queue (b) Acyclic queue.}
\vspace{-0.5cm}
\label{fig:topology}
\end{figure}

We consider multi-server queueing systems, with infinite queue size and First Come First Serve (FCFS) service discipline, as the building blocks of the service networks that we study. Furthermore, we study tandem queues and simple acyclic queueing networks as shown in Fig.\ref{fig:topology}. In a tandem topology, a customer must go through all the stages to receive the end-to-end service, while in an acyclic topology, the customers randomly go through one of the branches with the specified probabilities in Fig.~\ref{fig:topology}. Our MDN-based method does not assume a specific distribution for the service times or inter-arrival times and therefore, these processes can have non-stationary distributions. It should also be noted that our MDN-based method is not limited to the earlier mentioned assumptions about the service discipline or network topologies and we only consider them so that we can obtain theoretical baselines for comparison.

Consider a network consisting of $N$ queueing systems, where system $n$, $1\leq n \leq N$, is a multi-server queueing system with $c_n$ servers. Let $b_n^0$ denote the queue length (QL) of the $n$th queueing system upon arrival of the customer of interest. We define queue length information vector as $\bold{b} = [b_1^0, b_2^0, \cdots, b_{N}^0]$. Furthermore, $D(\bold{b})$ denotes the end-to-end delay of a new arrival given queue length information $\bold{b}$ upon arrival.
Our goal is to predict the distribution of a new arrival's end-to-end delay, based on the observed QL information upon arrival of the customer of interest.  In other words, we are interested in obtaining the distribution of $D(\bold{b})$. 
As we mentioned earlier, an important reason for estimating the distribution of the delay is to obtain probabilistic bounds instead of  making single-value predictions. More specifically, we can define probabilistic lower-bounds ($d_{lb}$) and upper-bounds ($d_{ub}$) as follows
\begin{align}
P(D(\bold{b}) > d_{ub}) &\leq \varepsilon_{ub}, \label{ub_bound}\\
P(D(\bold{b}) < d_{lb}) &\leq \varepsilon_{lb}, \label{lb_bound}
\end{align}
where $\varepsilon_{ub}$ and $\varepsilon_{lb}$ are the violation probabilities for the upper-bounds and the lower-bounds, respectively. Confidence interval is another statistic that will be used in this paper to measure the amount of uncertainty for each prediction. Since the confidence intervals will be used along with the MMSE predictions, we define the confidence interval for the random delay $D(\bold{b})$ as an interval with endpoints $ (E[D(\bold{b})]-x, E[D(\bold{b})]+ x)$ such that
\begin{equation}
P\Big(E[D(\bold{b})]-x < D(\bold{b}) < E[D(\bold{b})]+ x \Big) \geq P_{cl}, \label{eq:conf_int}
\end{equation}
where $P_{cl}$ denotes the corresponding confidence level.

Finally, we study single-value prediction of the end-to-end delay, which will be denoted by $\widehat{D}(\bold{b})$. In particular, we are interested in computing the Minimum Mean Square Error (MMSE) predictions, which can be obtained by the conditional expectation of the end-to-end delay, given QL information $\bold{b}$. In other words, it is well-known that $E[D(\bold{b})]$ minimizes the MSE of the predictor, which is defined as 
\begin{equation}
\text{MSE}(\widehat{D}(\bold{b})) \equiv E\left[\left(D(\bold{b}) - \widehat{D}(\bold{b})\right)^2 \right].
\end{equation}

\section{End-to-end Delay Analysis}\label{analysis}
In this section, we analyze the end-to-end delay of multi-stage queueing systems with tandem and acyclic topologies.  We only consider queue-length-based methods and investigate the differences between our approach and the existing ones. We begin with some approximations of the first two moments of the end-to-end delay. Using the first two moments, we discuss the normal approximation method, which motivates the use of mixtures of Gaussians for approximating the conditional distribution of the end-to-end delay.

\subsection{Analytical Expressions}\label{anal_exp}
We begin by considering a tandem network of $N$ queues as in Fig.~\ref{fig:tandem_topo}. 
We define the end-to-end delay of a customer as the sum of the waiting times and the service times that are experienced while going through the network, i.e.,
\begin{equation}
D = \sum_{n=1}^{N} \left(W_n+S_n \right), \label{eq:delay_ws}
\end{equation}
where $W_n$ and $S_n$ represent the waiting time and service time of the customer of interest at stage $n$. Let $b_n^\tau$, $\tau \in \{0,1,\cdots, N-1\}$, denote the queue length at stage $n$, once the customer of interest reaches stage $\tau+1$. As before, $b_n^0$ represents the queue length at stage $n$ once the customer of interest enters the network. In order to simplify the notation, we define  $q_n \equiv b_n^{n-1}$, which represents the queue length of the $n$'th queueing system upon arrival of the customer of interest at this stage.  In other words, $q_n$, $1 \leq n \leq N$, represent the sequence of queue lengths that the customer of interest observes as he goes through stages $1$ to $N$.  As will be discussed later in this section (see Fig.~\ref{fig:ql_update}), $q_n$ and $b_n^0$ are not necessarily equal for $n\geq2$ and therefore, the vector $\bold{q} = [q_1, q_2, \cdots, q_{N}]$ needs to be approximated from $\bold{b}$ and other system parameters. We propose an algorithm for this purpose at the end of this section. 
Now, having an estimation of $\bold{q}$, we can write $W_n$  as $\sum_{i=1}^{q_n+1} U_{n,i}$, where $U_{n,i}$, $1 \leq i \leq q_n+1$, denote the intervals between successive service completions at stage $n$. For mathematical tractability, we can approximate $U_{n,i}$ by $S_{n,i}/c_n$ under exponential service time assumption, where $S_{n,i}$ is a random variable with the same distribution as the service times in stage $n$. It should be noted that we only use the exponential service time assumption to make the derivations tractable, and we will not limit our analysis to a particular arrival or service time distribution. Now, using the approximation, we can write the end-to-end delay as
\begin{equation}
D(\bold{q}) \simeq_\mathcal{D} \sum_{n=1}^{N} \left( \sum_{i=1}^{q_n+1} \frac{S_{n,i}}{c_n}+S_n \right),\label{eq:delay}
\end{equation}
where '$\simeq_\mathcal{D}$' indicates equality (with approximation) in distribution. Assuming independent service times, the conditional mean and  variance of the end-to-end delay can be obtained from Eq.~(\ref{eq:delay}) as follows
\begin{align}
E[D(\bold{q})] & \simeq \sum_{n=1}^{N} \left( \frac{q_n+1}{c_n}+1  \right) E[S_n], \label{eq:mean_D} \\
Var[D(\bold{q})] & \simeq  \sum_{n=1}^{N} \left( \frac{q_n+1}{c_n^2}+1  \right) Var[S_n]. \label{eq:var_D}
\end{align}

As mentioned earlier, we are interested in the MMSE prediction, $E[D(\bold{b})]$, which can be approximated by $E[D(\bold{q})]$. The following Theorem states that using $E[D(\bold{q})]$ as our prediction of the end-to-end delay has the desirable property that the predictor becomes relatively more accurate, i.e., $Var[D(\bold{q})]/E[D(\bold{q})]^2\to0$, as the number of customers waiting in the queues or the number of stages in the network increases.

\begin{theorem} \label{theo:scv}
For a tandem queueing network with independent and exponentially distributed service times, we have
\begin{equation}
c_{D(\bold{q})}^2 = \frac{Var[D(\bold{q})]}{E[D(\bold{q})]^2} \to  0 \quad \text{as} \quad \sum_{n=1}^N (q_n+1) \to \infty,
\end{equation}
where $c_{D(\bold{q})}^2 $ is the squared coefficient of variation (SCV) of the end-to-end delay given $\bold{q}$.  
\end{theorem}
\begin{IEEEproof}
Let us define $Q = \sum_{n=1}^N (q_n+1)$. Using Eq.~(\ref{eq:var_D}), we have 
\begin{align}
Var[D(\bold{q})] &= \sum_{n=1}^N \frac{q_n+1}{c_n^2} Var[S_n] +  \sum_{n=1}^N  Var[S_n] \nonumber \\
  & \leq Q \beta_1 + \beta_2, \label{var_up}
\end{align}
where $\beta_1=\max_{1 \leq m \leq N}\{Var[S_m] /c_m^2\}$ and $\beta_2=\max_{1 \leq m \leq N} \{Var[S_{m}] \}$.
Similarly, using Eq.~(\ref{eq:mean_D}) we obtain
 \begin{align}
E[D(\bold{q})] &= \sum_{n=1}^N \frac{q_n+1}{c_n} E[S_n] +  \sum_{n=1}^N E[S_n] \nonumber \\
  & \geq Q \beta_3 + \beta_4, \label{mean_lo}
\end{align}
where $\beta_3 = \min_{1 \leq m \leq N}\{E[S_m] /c_m\}$ and $\beta_4 = \min_{1 \leq m \leq N} \{ E[S_{m}] \}$.
Combining Eqs.~(\ref{var_up}) and (\ref{mean_lo}), we have
\begin{equation}
c_{D(\bold{q})}^2 = \frac{Var[D(\bold{q})]}{E[D(\bold{q})]^2}  \leq \frac{ Q \beta_1 + \beta_2}{(Q \beta_3 + \beta_4)^2}.\label{csv_bound}
\end{equation}
Consequently, as $Q$ grows to infinity, $c_{D(\bold{q})}^2 < \beta_1/(Q \beta_3^2)$ and therefore, $c_{D(\bold{q})} \to 0$.

\end{IEEEproof}

\subsection{Updating Queue Lengths Through Time}\label{ql_update}

\begin{figure}[t!]
\centering
\includegraphics[angle=90,width=0.9\columnwidth]{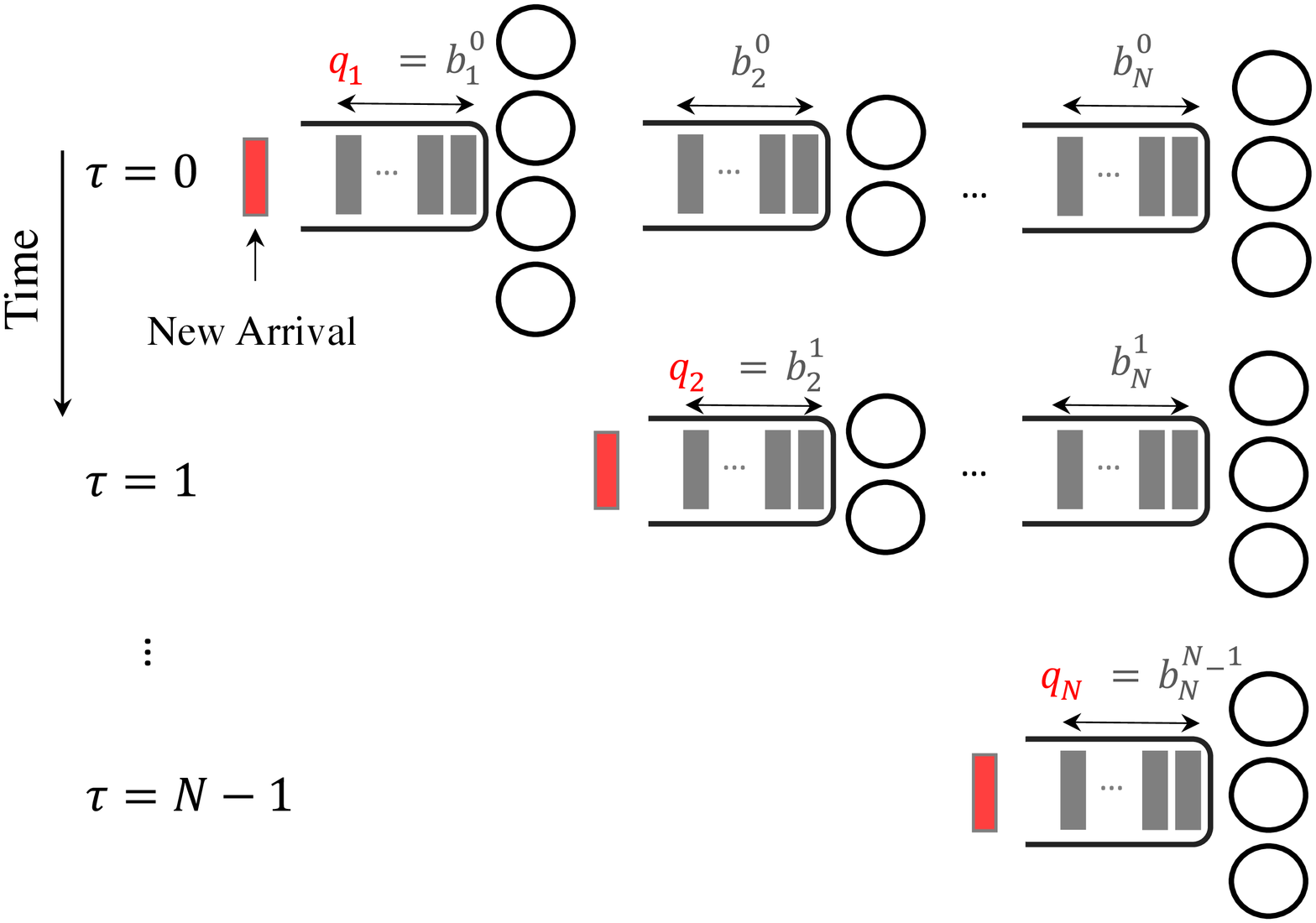}
\caption{Change of queue lengths as the customer of interest proceeds through a tandem network.}
\vspace{-0.5cm}
\label{fig:ql_update}
\end{figure}

In the previous subsection, we obtained our theoretical results based on the knowledge of queue lengths upon arrival of the customer of interest at each stage ($\bold{q}$). As shown in Fig.~\ref{fig:ql_update}, $\bold{q}$ is not necessarily equal to $\bold{b}$ and needs to be approximated. In this subsection, we explain an algorithm for approximating $\bold{q}$ from $\bold{b}$ in a tandem network, which is close to the proposed method in \cite{gue}.

Let us define $T_n$ as the experienced sojourn time of the customer of interest at stage $n$. Furthermore, $\mu_n$ represents the service rate of a single server at stage $n$. We consider $N-1$ updating steps, where the customer of interest proceeds from one stage to the next, in each step. 
The vector $\bold{q}$ is initialized to $\bold{b}$. In the first step, $\tau=0$, we update the number of customers in the down-stream stages, i.e., $2$ to $N$, given that the customer of interest has just arrived to the second queue. We assume that all $b_1^0+c_1$ customers in front of the customer of interest have arrived to the second stage. Furthermore, assuming busy servers during this time, we can estimate the number of departed customers from stage~$2$ by $c_2 \mu_2 E[T_1]$. Therefore, $b_2^1$ will be updated as $\max\{0, b_2^0 + b_1^0+c_1 - \lfloor{c_2 \mu_2 E[T_1]\rfloor}\}$. For stages $n > 2$, we update the queue lengths by $\max\{0, b_n^0 + \lfloor{c_{n-1} \mu_{n-1} E[T_1]\rfloor} - \lfloor{c_n \mu_n E[T_1]\rfloor}\}$, where the number of customers from the upstream stage are equal to $\lfloor{c_{n-1} \mu_{n-1} E[T_1]\rfloor}$. Similarly, in updating step $\tau > 1$, $b_n^\tau$, $\tau+1\leq n \leq N$, are updated as  explained in Algorithm~\ref{alg1}.

\begin{algorithm}
\caption{Calculating queue lengths upon arrival of the customer of interest at each stage}
\label{alg1}
\begin{algorithmic}

\STATE $q_1 \leftarrow b_1^0$
\FOR{$\tau=1$ to $N-1$} 
\STATE $b_{\tau+1}^\tau \leftarrow \max\{0, b_{\tau+1}^{\tau-1} + b_{\tau}^{\tau-1}+c_i - \lfloor{c_{i+1} \mu_{i+1} E[T_\tau]\rfloor}\}$ 
\STATE  $q_\tau \leftarrow b_{\tau+1}^\tau$
\FOR{$n=\tau+2$ to $N$}
\STATE $b_{n}^{\tau} \leftarrow \max\{0, b_{n}^{\tau-1} + \lfloor{c_{n-1} \mu_{n-1} E[T_\tau]\rfloor} - \lfloor{c_{n} \mu_n E[T_\tau]\rfloor}\}$ 
\ENDFOR
\ENDFOR
\end{algorithmic}
\end{algorithm}

A similar approach can be used for updating the queue lengths in an acyclic network, by taking into account the probabilities of each branch. It should be noted that this algorithm is based on heavy-traffic assumption and will only be used to obtain theoretical results for comparison with our main MDN-based method.

\section{Gaussian Mixture Model Approximation}\label{GMM}

In this section, we use our results from Section~\ref{anal_exp} to approximate the distribution of the end-to-end delay given queue length information $\bold{b}$. Since the end-to-end delay for a new customer consists of the intervals between service completion times of the customers ahead of the new arrival, the central limit theorem (CLT) and Eq.~(\ref{eq:delay}) suggest that the Normal distribution can be a good candidate for approximating the conditional distribution in a tandem network. Specifically, we approximate the conditional distribution of the delay in a tandem network with $D(\bold{b}) \sim \mathcal{N}(m(\bold{b}), \sigma^2(\bold{b}))$, where $m(\bold{b})$ and $\sigma^2(\bold{b})$ can be approximated by calculating $\bold{q}$ from $\bold{b}$ using Algorithm~\ref{alg1}, and then using Eqs.~(\ref{eq:mean_D}) and (\ref{eq:var_D}). Now, considering each path of the acyclic network as a tandem queue and using the normal approximation for each path, we can approximate the total distribution of the delay in an acyclic network by Gaussian mixture models (GMM), where the mixture weights are equal to the probabilities of taking each branch. More specifically, for an acyclic network as in Fig.~\ref{fig:acyclic_topo}, we have
\begin{equation}
P(D(\bold{b})) = \sum_{k \in \mathcal{P} } p_{k} \mathcal{N}(D|m_k(\bold{b}), \sigma_k^2(\bold{b})), \label{eq:mixture}
\end{equation}
where $\mathcal{P}$ is the set of existing paths in the acyclic network, and $m_k(\bold{b})$ and $\sigma_k^2(\bold{b})$ denote the mean and variance of the delay for path $k$, given queue length information $\bold{b}$ upon arrival. 
\begin{figure}[!t] 
\centering
\hspace{0.05cm}
\subfloat[]{\includegraphics[scale=0.35]{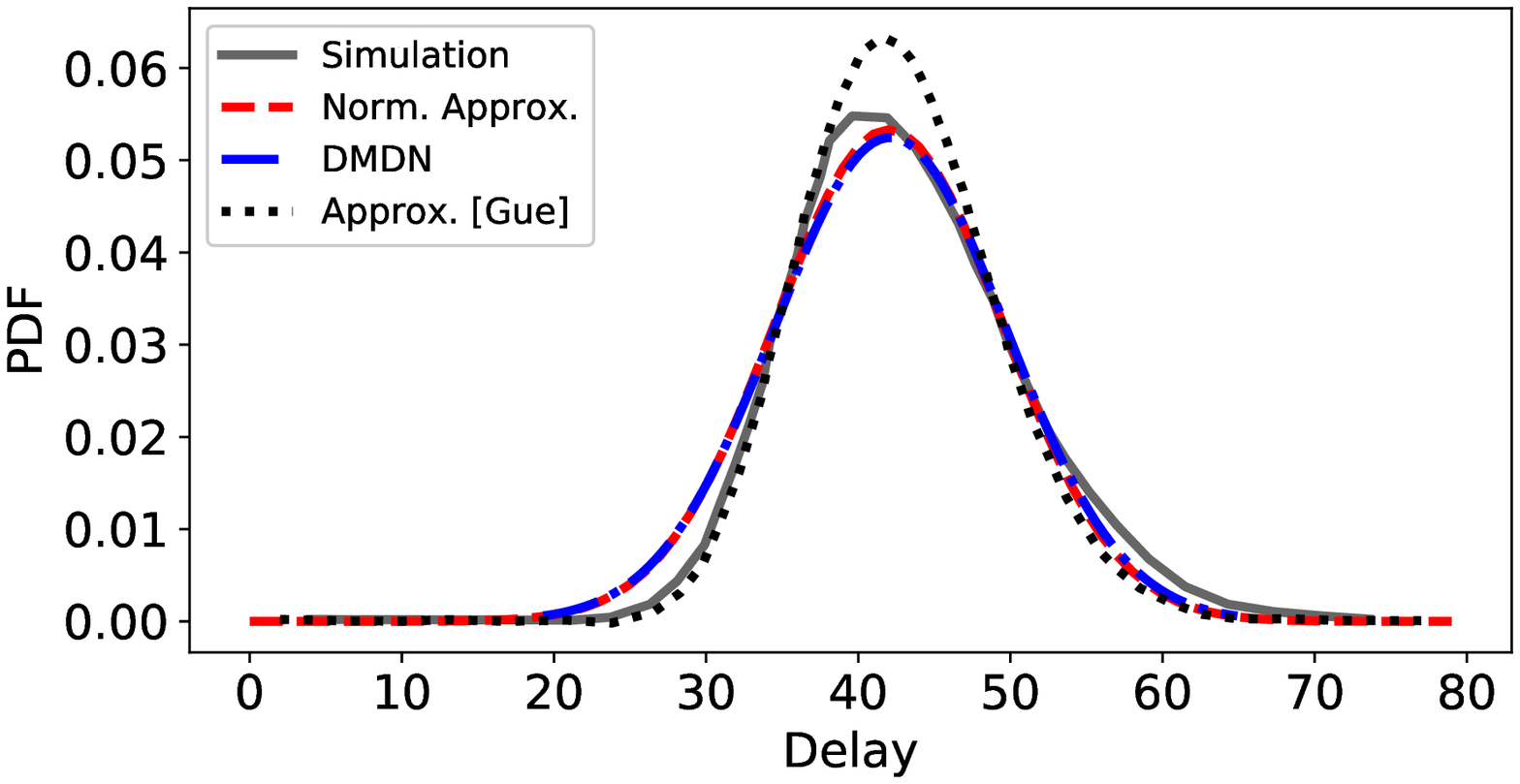}
\label{fig:PDF_comp_tandem}}
\\
\subfloat[]{\includegraphics[scale=0.35]{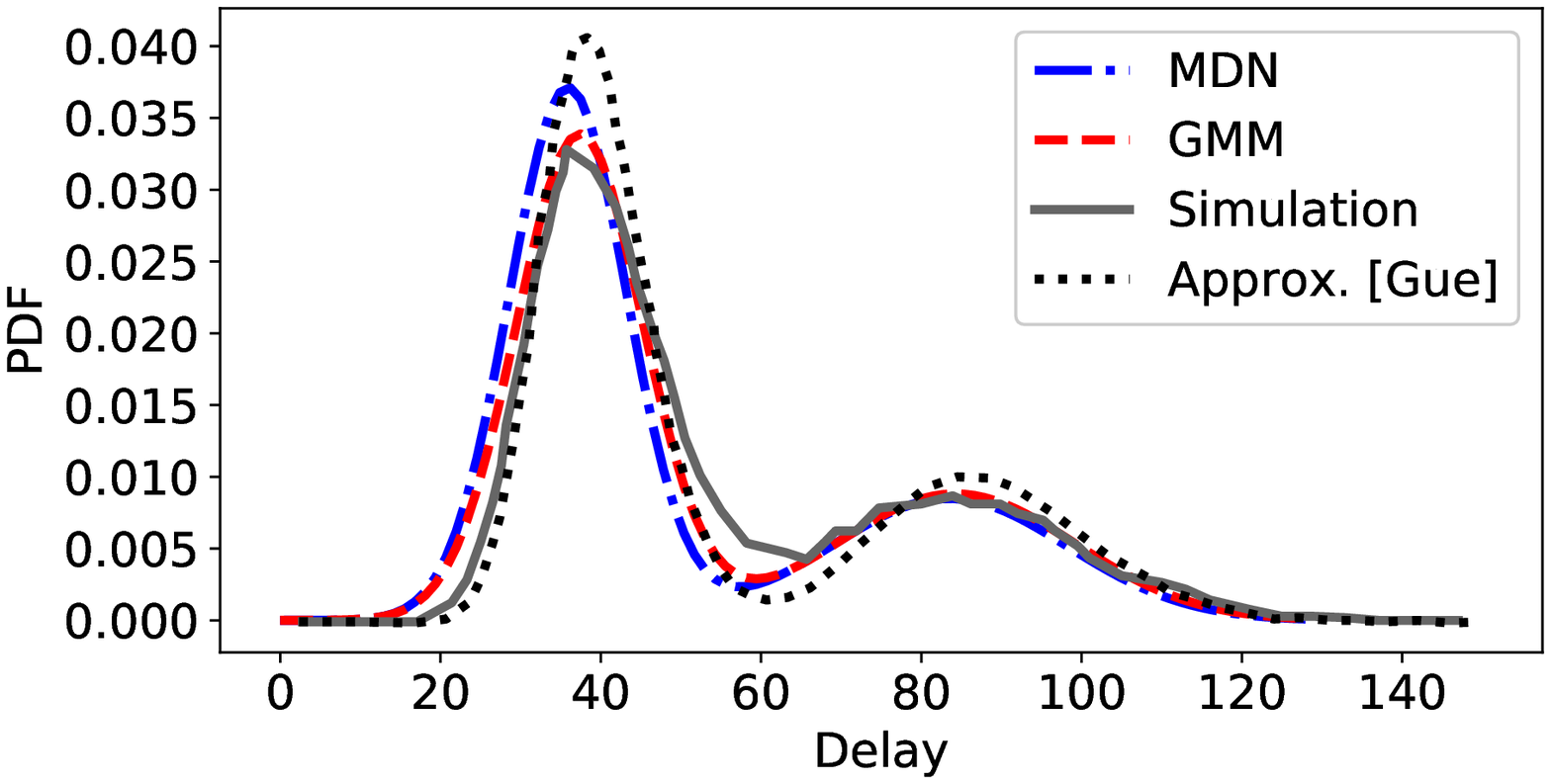}
\label{fig:PDF_comp_acyclic}}
\caption{Comparison of the conditional PDFs of the end-to-end delay in the a) tandem network, given queue lengths $\bold{b}=[6, 12, 13]$ b) acyclic network, given queue lengths $\bold{b}=[6, 4, 16, 13]$.}
\vspace{-0.5cm}
\label{fig:PDF_comp}
\end{figure}

In order to have a preliminary assessment of the proposed approximations, we perform some evaluations on similar network topologies as in~\cite{gue} (see Tandem~I and Acyclic~I topologies in Table~\ref{ta:topo_table}). Fig.~\ref{fig:PDF_comp} shows the comparison between the PDFs of the end-to-end delay obtained from the simulation, approximation method in~\cite{gue} and our GMM approximation method, for the tandem and acyclic networks. It can be observed that the normal distribution can be a good approximation of the conditional distribution of the delay for the tandem network. Furthermore, Fig.~\ref{fig:PDF_comp_acyclic} shows that the conditional distribution of the end-to-end delay of the acyclic network consists of two modes, since the conditional means of the two paths are far from each other relative to their standard deviations. As can be seen from Fig.~\ref{fig:PDF_comp}, the GMM approximation provides acceptable results, even in comparison with the more complex method of~\cite{gue}. However, both of these methods are limited to stationary systems with heavy traffic and require knowledge of the network topology, as well as other parameters of the network, such as the average service times, which might not be available in practice. In order to address these issues, we adopt a statistical learning approach to estimate the parameters of the Gaussian mixture models.
\subsection{Mixture Density Networks (MDNs)}\label{mdn}
As we discussed in the previous subsection, the Gaussian mixture model could be a good candidate for approximating the conditional distribution of the end-to-end delay in tandem or acyclic networks. However, the theoretical expressions obtained in Section~\ref{analysis} 
become less accurate as the number of customers in the network decreases or the degree of non-stationarity increases. Nevertheless, we can still approximate the conditional distribution of the end-to-end delay by GMMs, since they are powerful enough to approximate arbitrary distributions~\cite{bishop}. 

\begin{figure}[t!]
\centering
\includegraphics[angle=90,width=0.9\columnwidth]{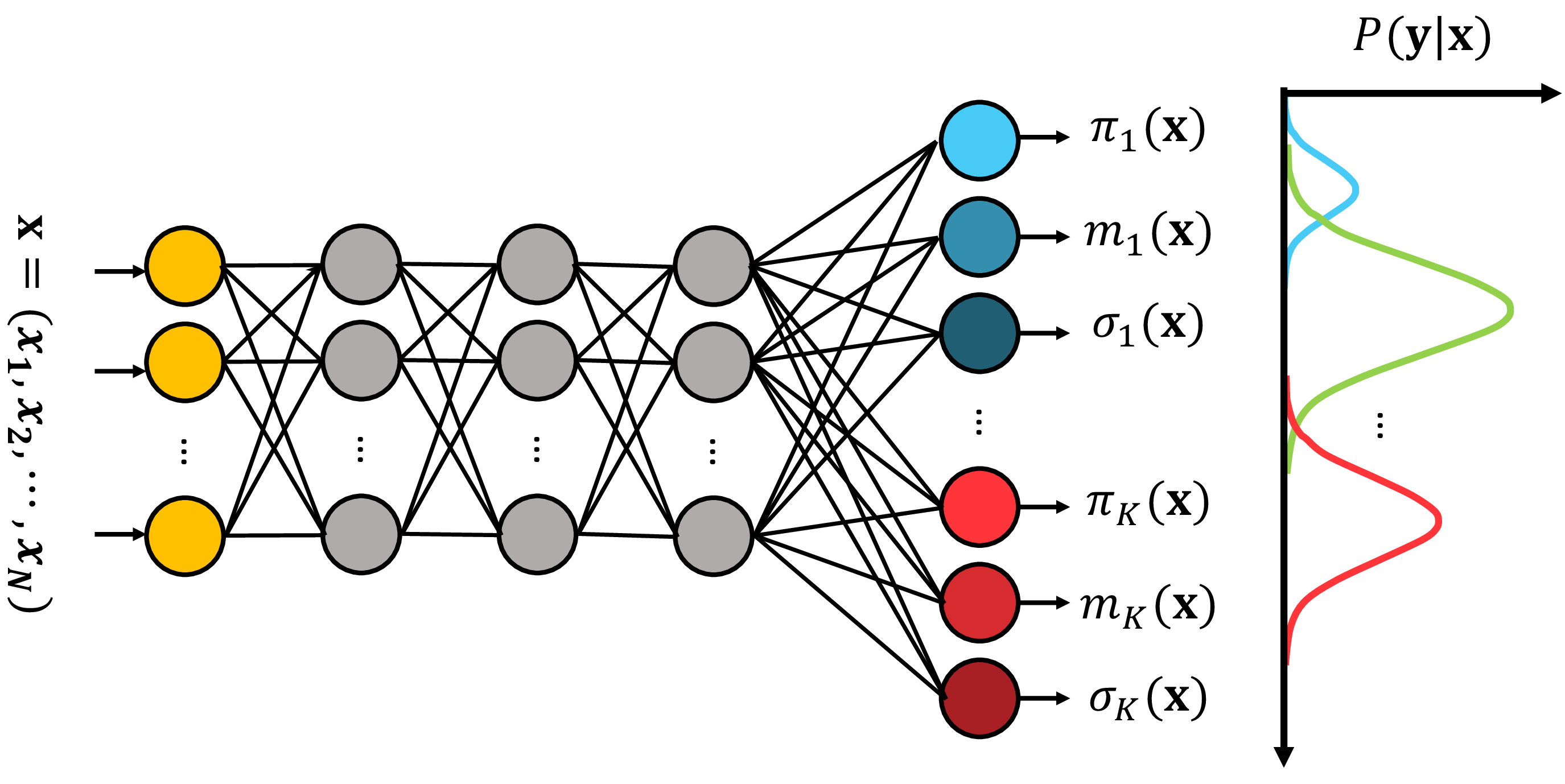}
\caption{Using mixture density network (MDN) for approximating the conditional distribution of $\bold{y}$ given $\bold{x}$.}
\vspace{-0.4cm}
\label{fig:mdn}
\end{figure}

In order to address the problems related to GMM parameter estimation under more realistic assumptions, we adopt a statistical learning approach called \emph{Mixture Density Networks} (MDN). 
The MDN provides a general framework for approximating arbitrary conditional distributions using mixture models.  Considering Gaussian components, an MDN approximates the conditional distribution of $\bold{y}$ given $\bold{x}$ by
\begin{equation}
P(\bold{y}|\bold{x}) = \sum_{k=1}^K \pi_k(\bold{x}) \mathcal{N}(\bold{y}|m_k(\bold{x}), \sigma_k^2(\bold{x})), \label{eq:mdn}
\end{equation}
where $\pi_k(\bold{x}) \in (0, 1)$ are the mixing coefficients and,  $m_k(\bold{x})$ and $\sigma_k^2(\bold{x})$ denote the mean and variance of the $k$'th kernel, $0 \leq k \leq K$, given  $\bold{x}$. 
An MDN estimates the parameters of the mixture model using a fully-connected neural network. As shown in Fig.~\ref{fig:mdn},  the output layer consists of three types of nodes which predict the parameters of the mixture model in Eq.~(\ref{eq:mdn}). The first type uses the soft-max activation function to predict the mixing coefficients such that $0 \leq \pi_k \leq 1$ and $\sum_k \pi_k =1$. The second group, which predict the variances of the kernels, use  exponential activations to ensure non-negative values. The last group of the nodes use linear activations and compute the means of the kernels. Using a data set of $N_{sample}$ observations (queue lengths) and their corresponding target values (end-to-end delay), $\{(\bold{x}_j=\bold{b}_j, \bold{y}_j = D_j)|1 \leq j \leq N_{sample}\}$, the mixture density network learns the weights of the neural network by minimizing the error function, which is defined as the negative logarithm of the likelihood, i.e.,
\begin{equation}
E = - \sum_{j =1}^{N_{sample}} \ln \left\{ P(\bold{y}_j|\bold{x}_j)\right\}.
\end{equation}
We refer the reader to~\cite{speech, world_mod} for more information on the applications of the MDNs.

\section{Evaluation and Numerical Results}\label{eval}
In this section, we evaluate the accuracy of our proposed methods and compare some of our results to the existing method from~\cite{gue} under different network topologies that are summarized in Table~\ref{ta:topo_table}. Furthermore, we consider multiple types of arrival processes, including non-stationary and non-renewal arrivals (Table~\ref{ta:sim_table}). It should be noted that time is normalized by the mean service time of the ingress queue, i.e., $1/(c_1 \mu_1)$, in all of the following experiments.

Let us start with revisiting the delay distribution prediction experiment in Section~\ref{GMM} and compare our MDN-based predictions to the previously obtained theoretical and simulation results. Our MDN-based predictor consists of three hidden layers with 64, 32 and 32 hidden nodes, and uses RELU activation function for the hidden layers. The MDN layer outputs the parameters of a Gaussian mixture model with 3 kernels. The queue lengths of all the stages in the network will be used as the feature set, while the ground-truth end-to-end delays serve as the labels. The MDN-based predictor is trained for 500 epochs with a batch size of 512. Fig.~\ref{fig:PDF_comp} shows the predicted probability density function obtained from the MDN-based method along with the simulation and theoretical results, for the Tandem~I and Acyclic~I topologies (see Table~\ref{ta:topo_table}). As we can see, the MDN can provide acceptable estimations of the conditional distribution, without any knowledge of the network topology or its parameters. Furthermore, once the neural network has been trained, the distribution predictions can be obtained in real-time. This can be a huge benefit compared to the more complex methods, such as the approximation method in~\cite{gue}, which require a large number of convolution operations to perform.  

\begin{table}[t!]
\caption{Network Topologies.}
\vspace{-0.5cm}
\centering
\begin{tabular}[t]{llll}
\hline
\bfseries Topology& \bfseries \parbox{5mm}{Num.~of~servers\\$[c_1,c_2, \cdots, c_N]$} & \bfseries\parbox{5mm}{Service~rates\\$[\mu_1,\mu_2, \cdots, \mu_N]$}&\bfseries\parbox{5mm}{Arrival\\type}\\ 
\hline
\hline
Tandem I&$[5,3,2]$&$[0.2,0.33,0.5]$&Gamma\\ 
Tandem II & $[3,3,5,5,4,4]$&$[0.33,0.33, \cdots,0.33]$&NHPP\\ 
Acyclic I&$[5,3,3,2]$&$[0.2,0.22,0.11,0.5]$&Gamma \\ 
Acyclic II&$[1,1,1,1,1]$&$[1.0, 0.44, 0.33, 0.22, 1.0]$&MMPP \\ 
\end{tabular}\label{ta:topo_table}
\end{table}


\begin{table}[t!]
\caption{Arrival and service models.}
\vspace{-0.5cm}
\centering
\begin{tabular}[t]{ll}
\hline
\bfseries Distribution& \bfseries Parameters \\ 
\hline
\hline
Gamma (Arrival)&$\lambda = 0.95$, SCV = $0.7$\\ 
NHPP (Arrival)&$\bar{\lambda} = 0.95$, $\alpha=0.5$, $T_p=144$\\ 
MMPP (Arrival)&$\bar{\lambda} = 0.95$, $P_{on\to off}$ = 0.4,  $P_{off\to on}$ = 0.1\\ 
\hline
Gamma (Service time)& for $\mu$ see Table~\ref{ta:topo_table}, SCV = $0.8$\\ 
\end{tabular}\label{ta:sim_table}
\vspace{-0.5cm}
\end{table}

\subsection*{Application to Service Function Chaining}
\begin{figure}[!t] 
\centering
\subfloat[]{\includegraphics[scale=0.3]{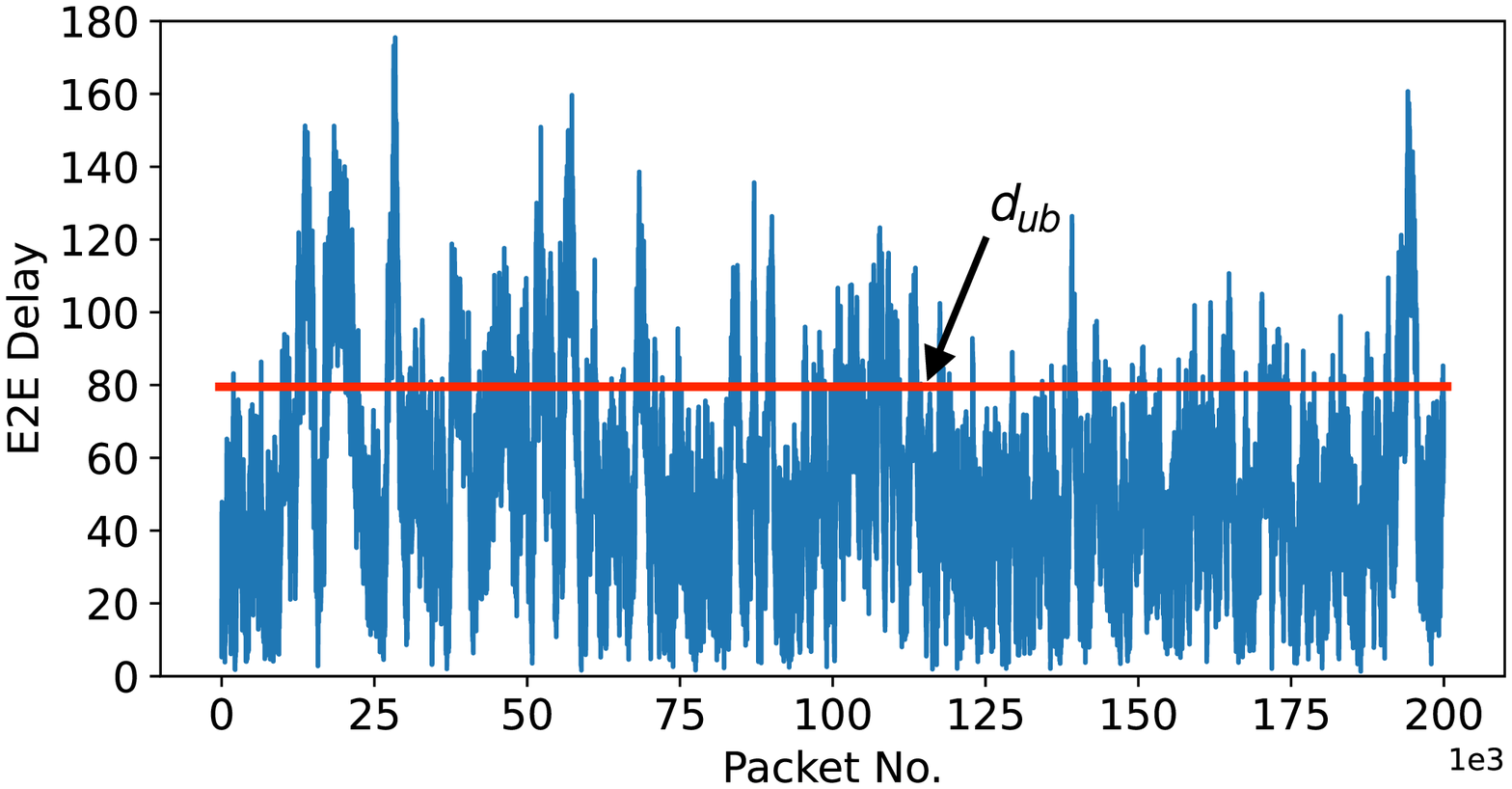}
\label{fig:tandem_without_AC}}
\\
\subfloat[]{\includegraphics[scale=0.3]{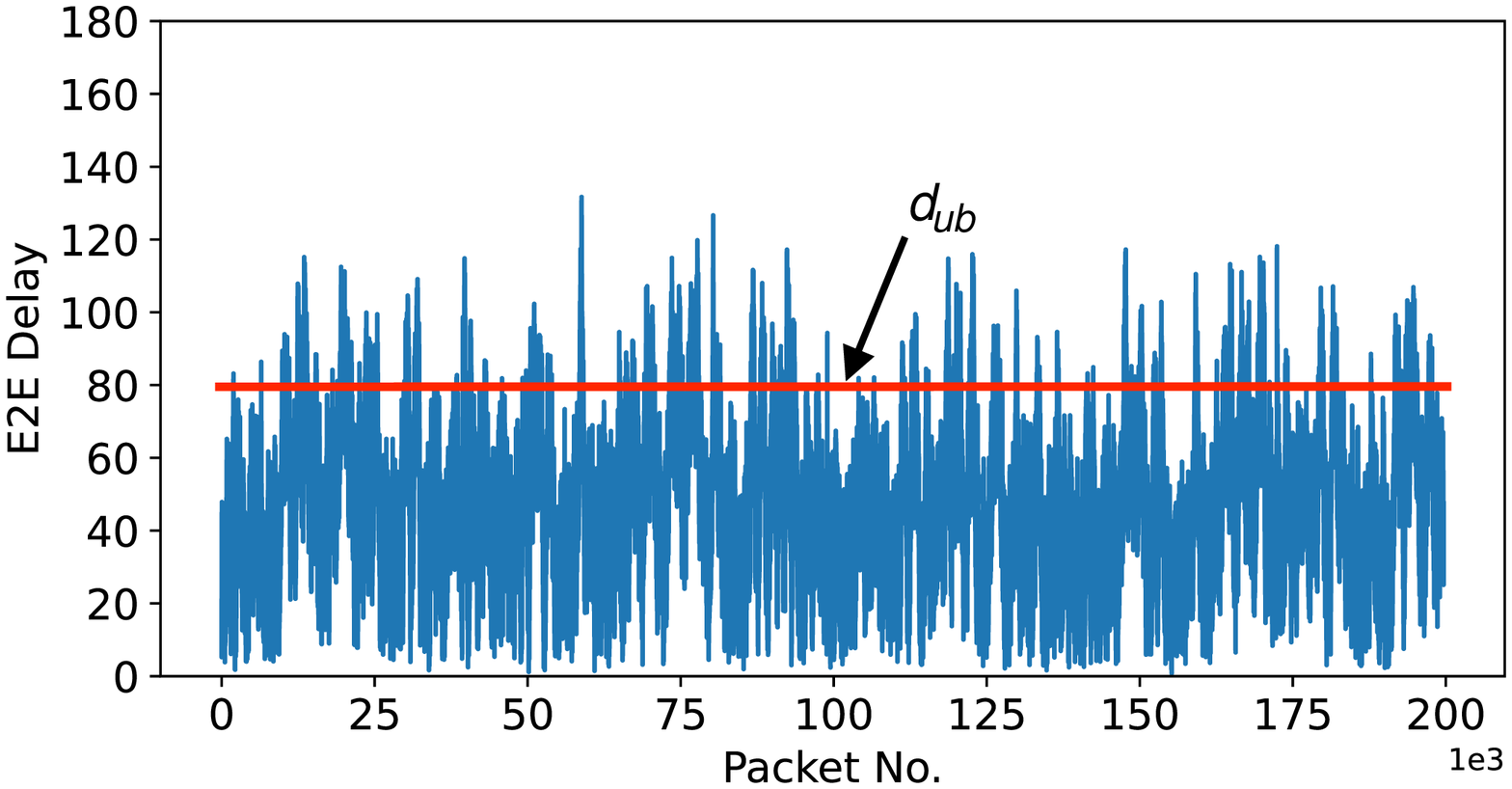}
\label{fig:tandem_with_AC}}
\caption{End-to-end packet delay in a service chain: a) without admission control b) with admission controller that guarantees $P(D>d_{ub})<0.05$}
\label{fig:AC}
\end{figure}

\begin{figure}[!t] 
\centering
\subfloat[]{\includegraphics[angle=90,width=0.4\columnwidth]{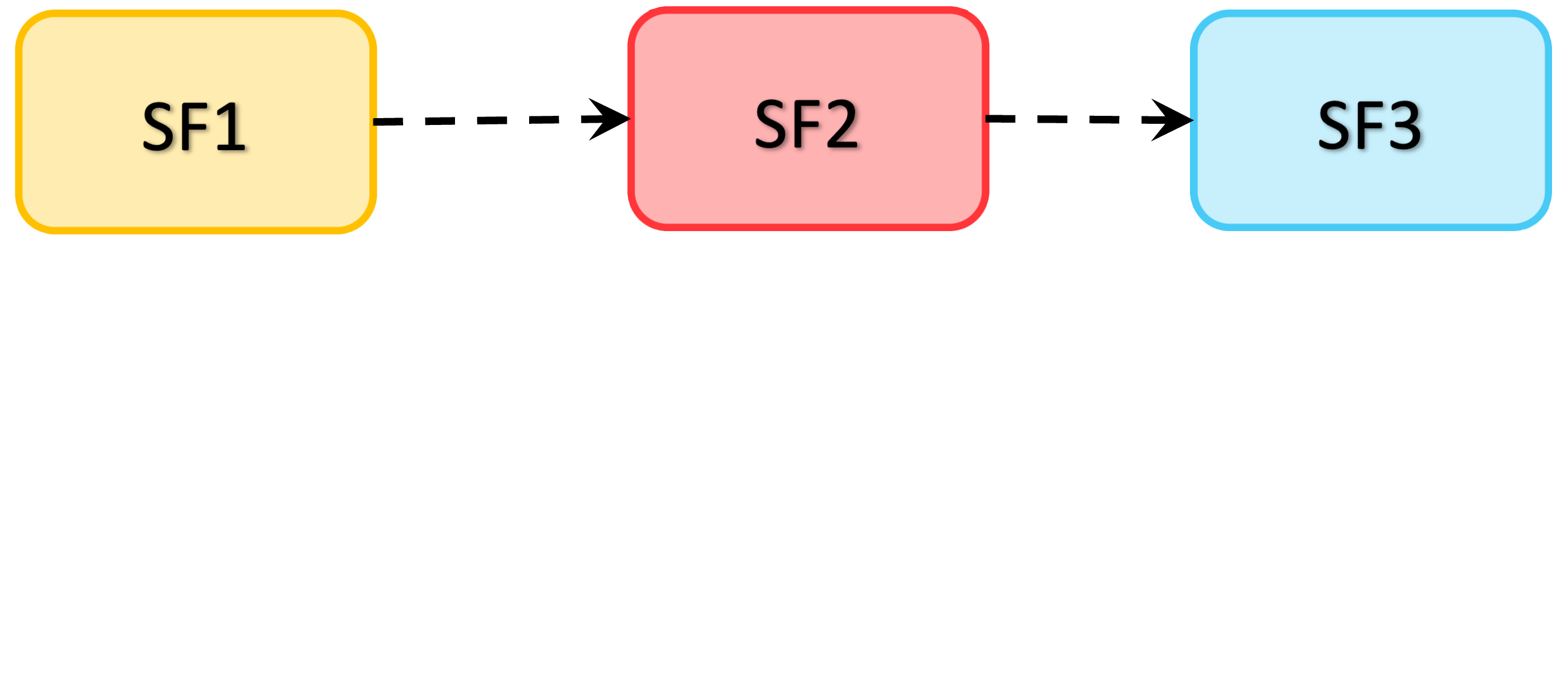}
\label{fig:sfc_graph}}
\quad
\subfloat[]{\includegraphics[angle=90,width=0.4\columnwidth]{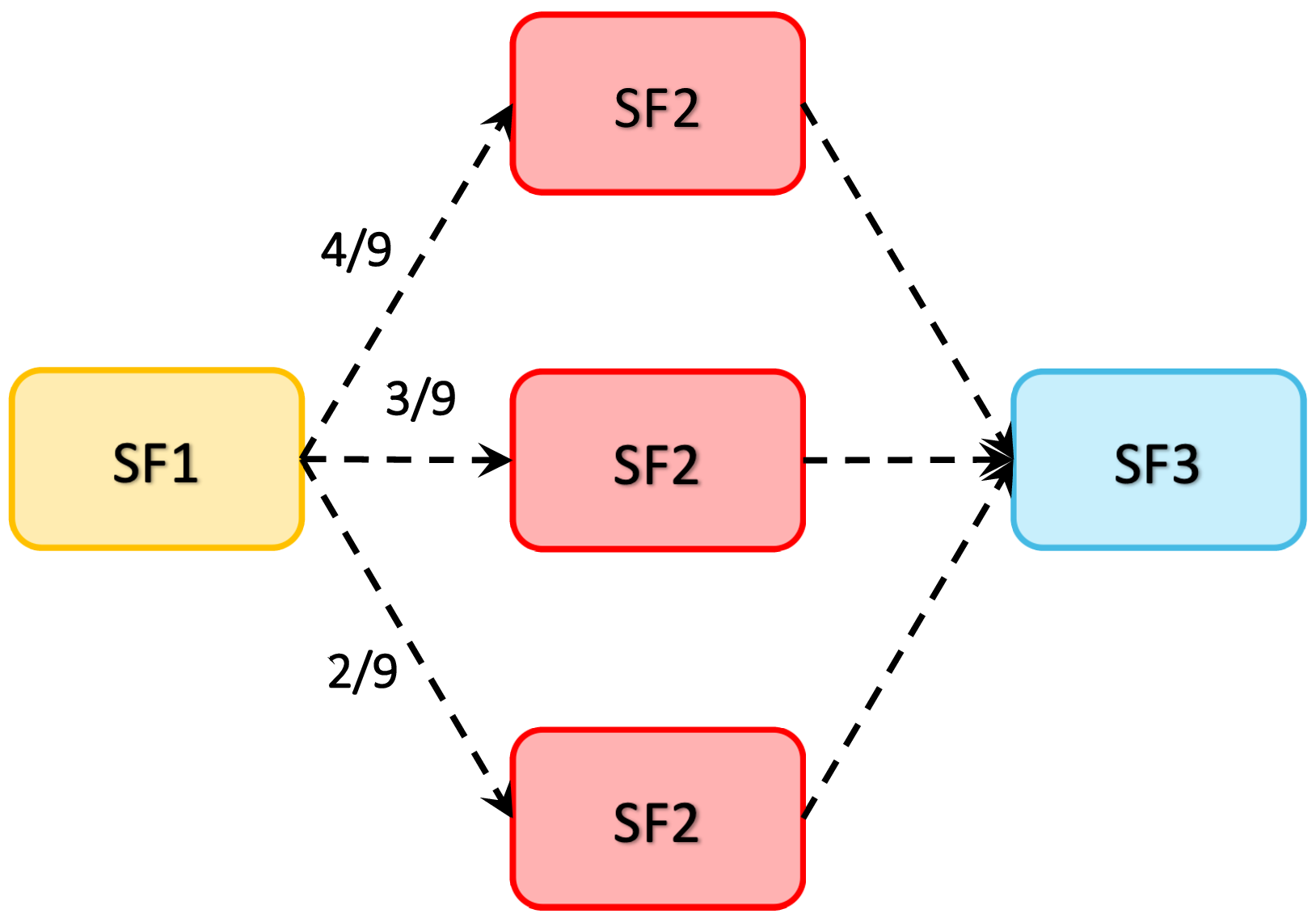}
\label{fig:sfc_real_topo}}
\caption{Service function chain: a) tandem b) acyclic.}
\vspace{-0.5cm}
\label{fig:sfc}
\end{figure}

As discussed in Section~\ref{intro}, service function chaining is one of the examples of the service networks that can benefit from our MDN-based method by providing real-time service guarantees. We model each service function with a multi-server queueing system, where each server represents an instance of a service function.
\\\\
\noindent \textbf{Admission Control:} Let us consider a service function chain as in Fig.~\ref{fig:sfc}a, which is modeled by a queueing network with topology Tandem~I. Moreover, we assume that every packet must be processed within an end-to-end deadline of $d_{ub}=80$, otherwise it is considered useless. Our goal is to design an admission controller that drops the packets with high probability of missing the deadline (more than $95\%$)  at the entrance of the service chain, so that there will be more room for the following packets and therefore, higher throughputs can be obtained. Fig.~\ref{fig:AC} shows the end-to-end delay of around $200$K packets that have made it through the service chain, both with and without the admission controller. As can be observed, smaller number of packets will miss the deadline with admission control (Fig.~\ref{fig:AC}b), which shows how dropping a few packets that have high probability of missing the deadline at the entrance can have large impacts on the end-to-end delay of  the following packets. Although the admission controller only rejects the packets for which $P(D>d_{ub}) \geq 0.95$, the end-to-end throughput has been increased from $87\%$ to $95\%$. The real-time prediction of the delay distribution in a SFC can be used for different purposes such as SLA compliance prediction and scaling service function chains.
\\

\begin{figure}[t!]
\centering
\hspace{-0.7cm}
\includegraphics[scale=0.35]{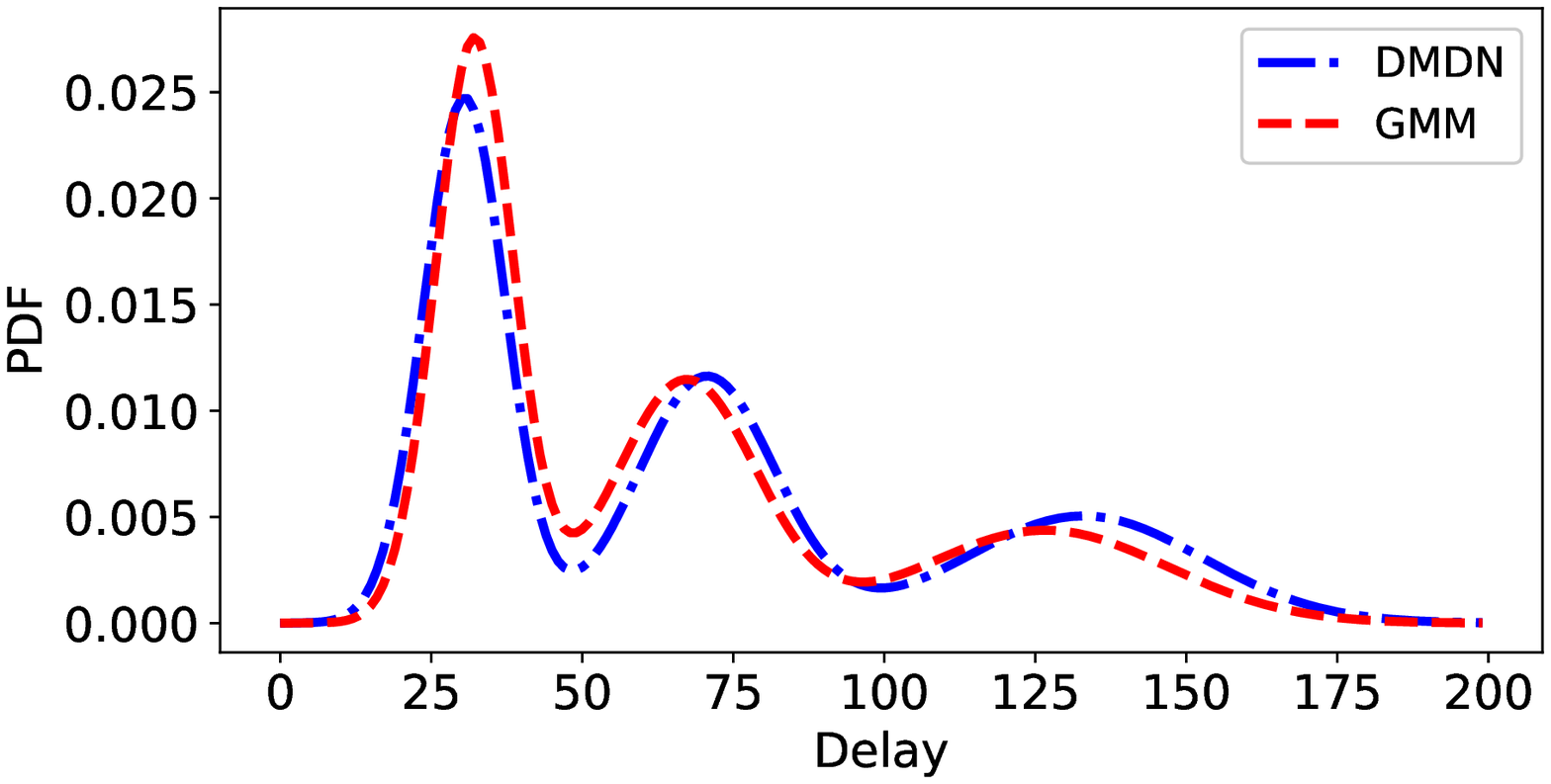}
\caption{Comparison of the conditional PDFs of the end-to-end delay in the SFC shown in Fig.~\ref{fig:sfc}b, given queue lengths $\bold{b}=[6, 5, 15, 20, 10]$. }
\vspace{-0.5cm}
\label{fig:pdf_sfc}
\end{figure}

\begin{figure}[!t] 
\centering
\subfloat[]{\includegraphics[scale=0.35]{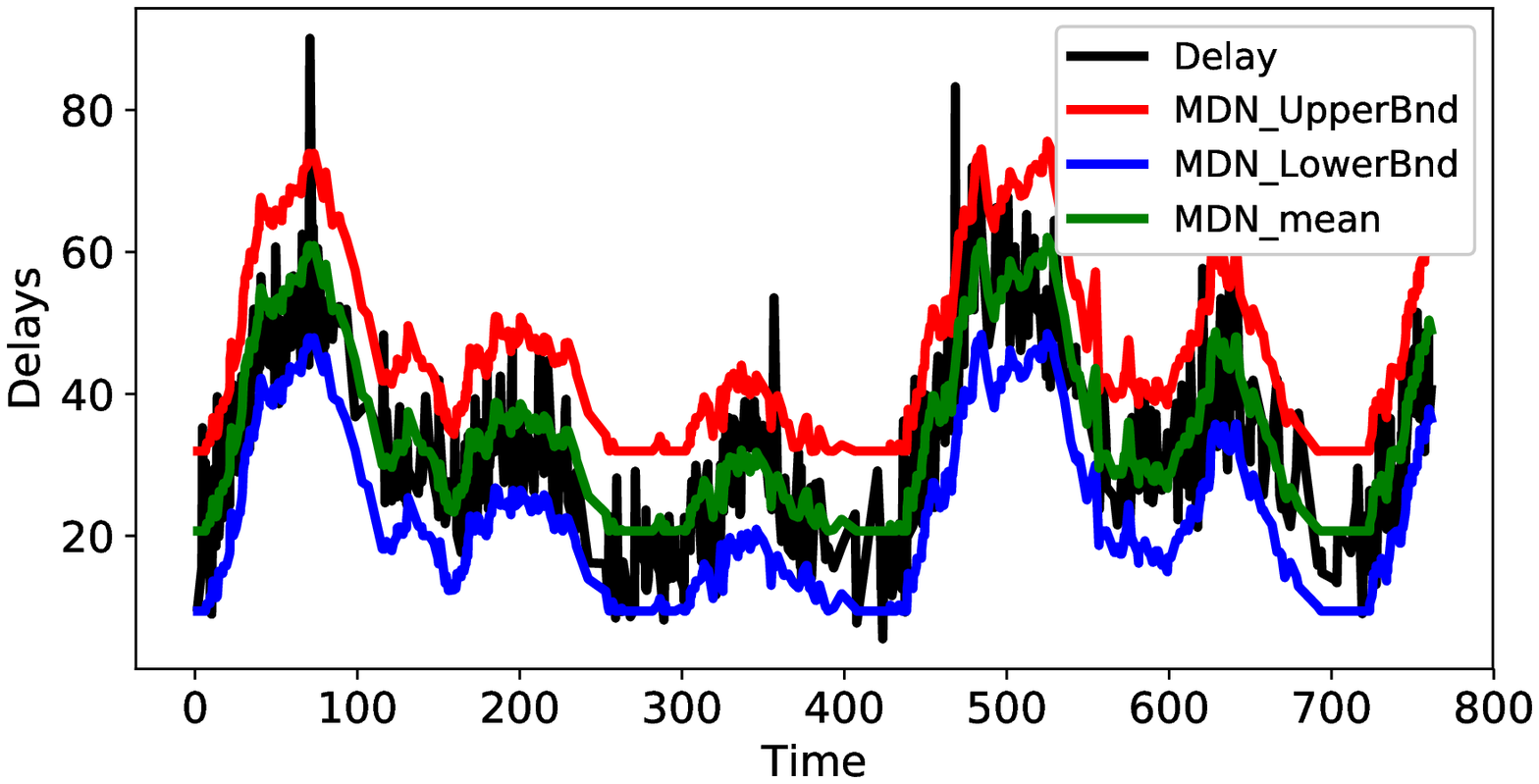}
\label{fig:ql_bounds_nhpp}}
\\
\subfloat[]{\includegraphics[scale=0.35]{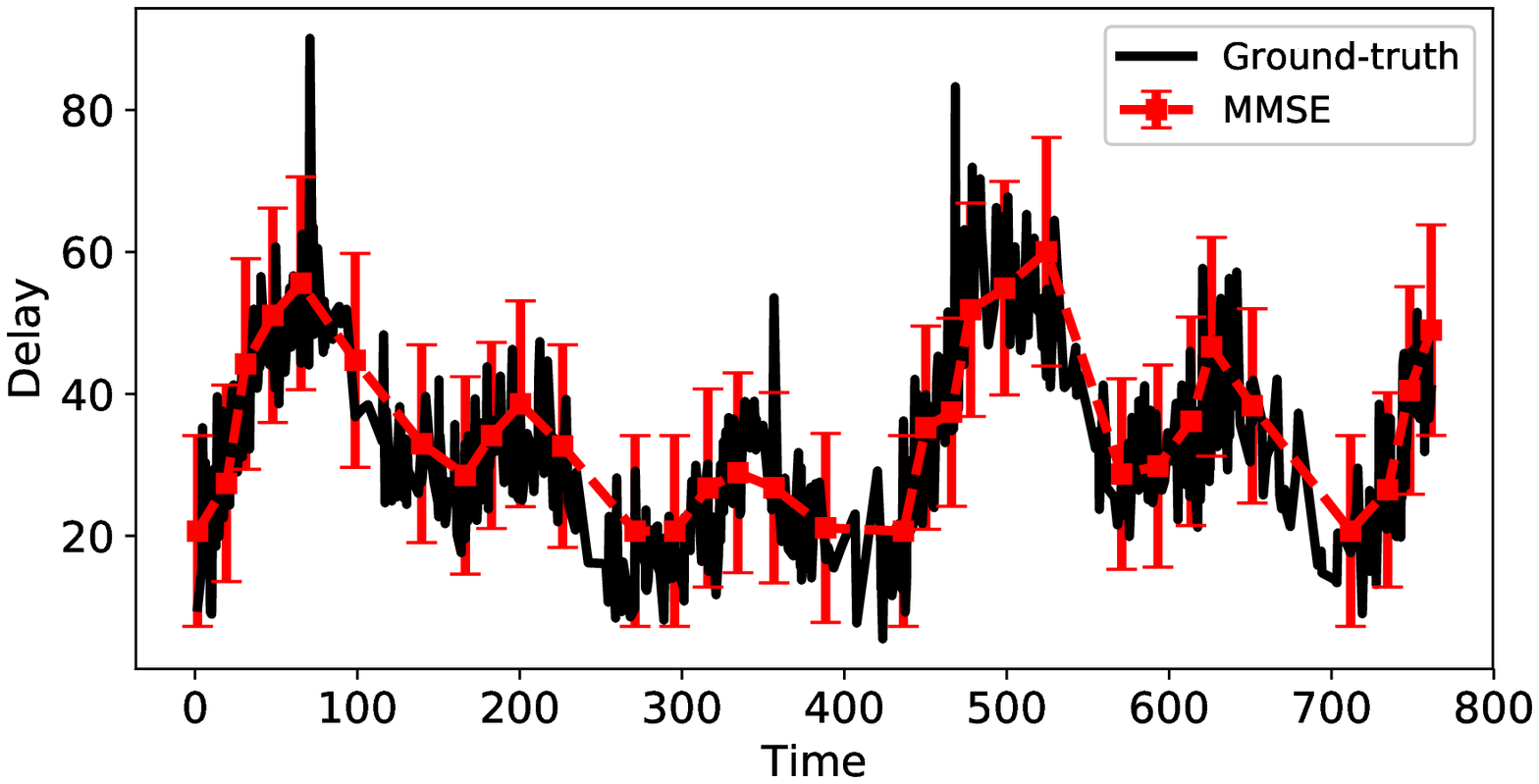}
\label{fig:conf_bounds_nhpp}}
\caption{A sample-path of the end-to-end delay with a) probabilistic upper bounds and lower bounds with violation probabilities $\varepsilon_{lb}=\varepsilon_{ub}=0.05$. b) MMSE predictions and their corresponding 95\% confidence intervals.}
\vspace{-0.5cm}
\end{figure}

\noindent \textbf{Multimodal Distribution:}  Consider a service chain comprising of three service functions as in Fig.~\ref{fig:sfc}b.  The incoming traffic to SF~1 will be forwarded to one of the three instances of SF~2 by probabilities $4/9$, $1/3$ and $2/9$ (Fig.~\ref{fig:sfc}b). Finally, all instances of SF~2 forward their packets to SF~3. This service chain can be modeled by the acyclic queueing network in Fig.\ref{fig:acyclic_topo}, parameters of which are summarized in Table~\ref{ta:topo_table} (topology Acyclic~II). Moreover, we assume \emph{non-renewal} arrivals, which are modeled by Markov Modulated Poisson Process (MMPP) as described in Table~\ref{ta:sim_table}. Fig.~\ref{fig:pdf_sfc} shows the comparison between the PDFs obtained from the GMM approximation method and the MDN-based method, given queue lengths $\bold{b}=[6, 5, 15, 20, 10]$. As can be observed, the MDN-based predictor is capable of capturing the three existing modes in the conditional distribution of the delay, and has a good match with the GMM approximation method. 
\\\\
\noindent \textbf{Probabilistic Bounds:} As we discussed earlier, the predicted distributions can also be used to obtain probabilistic bounds on the end-to-end delay. For this experiment, we consider a service chain consisting of $5$ service functions in tandem, parameters of which are summarized in Table~\ref{ta:topo_table} (topology Tandem~II). Furthermore, we assume \emph{non-stationary} arrivals modeled by Non-homogeneous Poisson Process (NHPP). We adopt the same model as in~\cite{ibrahim_thesis} with sinusoidal arrival rate, i.e., we consider an arrival rate of $\lambda(t) = \bar{\lambda}(1+\alpha \sin(2 \pi t/T_p))$, where  $\bar{\lambda}$, $\alpha$ and $T$ represent the average arrival rate, relative amplitude and the cycle length of the arrival rate. These parameters are summarized in Table~\ref{ta:sim_table}. Fig.~\ref{fig:ql_bounds_nhpp} shows the sample paths of the actual end-to-end delay along with the probabilistic upper bounds, lower bounds and the conditional mean, obtained from the learned distribution. The bounds are computed for violation probabilities $\varepsilon_{lb}=\varepsilon_{ub}=0.05$. It should be noted that there exists a trade off between the tightness of the bounds and the violation probabilities. 
Similarly, Fig.~\ref{fig:conf_bounds_nhpp} shows the MMSE predictions and the $95\%$ confidence intervals computed from Eq.~(\ref{eq:conf_int}). As can be observed, using the confidence intervals along with the predictions, which are shown by the error bars, can be much more informative compared to the single value predictions, since it provides a region in which the ground-truth delays are more likely to occur.

\section{Conclusions}\label{con}
In this paper, we attempted to show the potential of the statistical learning methods in providing real-time probabilistic bounds on the end-to-end delay of  service systems. In particular, we studied the problem of delay prediction and distribution estimation in multi-stage queueing networks, based on queue length information. We showed that our MDN-based method, which only uses the queue length information, can be used to predict the conditional distribution of the end-to-end delay, without requiring any knowledge of the system parameters or network topology. Furthermore, the estimated distribution can be used to obtain much more informative statistics, such as probabilistic bounds, compared to the common single-value predictions. 
It would be interesting to extend this work by exploiting the MDN-based method in the design of a controller for VNF chains, where the admission control or auto-scaling is done based on the predicted delay distribution.

\vspace{12pt}

\end{document}